\documentclass[twocolumn,amsmath,amssymb,prb,groupedaddress,fleqn]{revtex4-2}
\usepackage{epsfig}
\usepackage{graphicx}
\usepackage{dcolumn}
\usepackage{bm}
\usepackage{color} 

\usepackage{ulem}
\usepackage{amsmath}

\usepackage{here}

\usepackage[bookmarks=false,hyperfigures=true,linkcolor=blue,urlcolor=blue,citecolor=blue,colorlinks=true]{hyperref}

\begin{document}


\title{Spin-Orbit Torque Switching of Noncollinear Antiferromagnetic Antiperovskite Manganese Nitride Mn$_3$GaN}

\author{T. Hajiri}
 \email[Electronic mail:~]{t.hajiri@nagoya-u.jp}
\affiliation{Department of Materials Physics, Nagoya University, Nagoya 464-8603, Japan}
\author{K. Matsuura}
\affiliation{Department of Materials Physics, Nagoya University, Nagoya 464-8603, Japan}
\author{K. Sonoda}
\affiliation{Department of Materials Physics, Nagoya University, Nagoya 464-8603, Japan}
\author{E. Tanaka}
\affiliation{Department of Materials Physics, Nagoya University, Nagoya 464-8603, Japan}
\author{K. Ueda}
\affiliation{Department of Materials Physics, Nagoya University, Nagoya 464-8603, Japan}
\author{H. Asano}
\affiliation{Department of Materials Physics, Nagoya University, Nagoya 464-8603, Japan}
\date{\today}
%
\begin{abstract}
Noncollinear antiferromagnets have promising potential to replace ferromagnets in the field of spintronics as high-density devices with ultrafast operation.
To take full advantage of noncollinear antiferromagnets in spintronics applications, it is important to achieve efficient manipulation of noncollinear antiferromagnetic spin.
Here, using the anomalous Hall effect as an electrical signal of the triangular magnetic configuration, spin--orbit torque switching with no external magnetic field is demonstrated in noncollinear antiferromagnetic antiperovskite manganese nitride Mn$_3$GaN at room temperature.
The pulse-width dependence and subsequent relaxation of Hall signal behavior indicate that the spin--orbit torque plays a more important role than the thermal contribution due to pulse injection.
In addition, multistate memristive switching with respect to pulse current density was observed.
The findings advance the effective control of noncollinear antiferromagnetic spin, facilitating the use of such materials in antiferromagnetic spintronics and neuromorphic computing applications.
\end{abstract}


\maketitle
\section{INTRODUCTION}
Noncollinear antiferromagnetic (AFM) materials have attracted significant attention in basic and applied science because in addition to having the excellent properties of collinear AFM---such as fast dynamics, suitability for high-density integration, and stability against external perturbations---noncollinear AFM materials can overcome the weakness of collinear AFM materials---namely the small electrical signal---by the anomalous Hall effect (AHE)~\cite{Mn3Sn AHE, AFM spintronics1, AFM spintronics2, AFM spintronics3}.
Thus, efficient control of noncollinear AFM spin is essential for the application of such materials in AFM spintronics.
In the past decade, electrical manipulation via spin--transfer torque (STT) and spin--orbit torques (SOTs) has become one of the most promising techniques in the field of ferromagnet (FM)--based spintronics, not only to take the place of dynamic random access memory in the current computer memory hierarchy~\cite{STT SOT MRAM} but also to instantiate multistate magnetoresistive random access memories for neuromorphic computing~\cite{neuromorphic1, neuromorphic2}.

Recent studies demonstrate that like collinear AFM materials, noncollinear AFM materials can be controlled via SOT in the same way as FM materials.
Collinear AFM materials consisting of NiO/Pt bilayers and Pt/NiO/Pt trilayers show a critical current density ($J_\mathrm{c}$) on the order of 10$^7$–-10$^8$~A/cm$^{2}$~\cite{SOT_NiO1, SOT_NiO2, SOT_NiO3}, similar to that in typical FM/heavy metal (HM) bilayers.
In contrast, $J_\mathrm{c}$ values one to two orders of magnitude smaller $J_\mathrm{c}$ have been observed in noncollinear AFM Mn$_3$GaN (MGN)/Pt bilayers ($1.5\times10^6$~A/cm$^{2}$)~\cite{Hajiri APL} and Mn$_3$Sn/W bilayers ($5\times10^6$~A/cm$^{2})$~\cite{Mn3Sn SOT}.
In the case of a collinear AFM system, 90$^{\circ}$ switching of the N$\rm{\acute{e}}$el vector is required because the electrical signal, such as spin Hall magnetoresistance and anisotropic magnetoresistance, is maximal.
To achieve 90$^{\circ}$ switching of the N$\rm{\acute{e}}$el vector, diagonal current flow using an eight- or four-terminal device with complex electrical write/read operation is required~\cite{SOT_NiO1, SOT_NiO2, SOT_NiO3}, except for CuMnAs using two terminal writing device~\cite{CuMnAs_2terminal}.
A noncollinear AFM system, by contrast, needs 180$^{\circ}$ switching of each spin of triangular magnetic configuration, which can be accomplished using a simple four-terminal Hall device with no external magnetic field in Mn$_3$GaN~\cite{Hajiri APL}, or with an external magnetic field in Mn$_3$Sn~\cite{Mn3Sn SOT}.
Therefore, in addition to their large electrical signals due to non-zero Berry curvature~\cite{Chen_PRL}, noncollinear AFM systems have a distinct advantage in spintronic applications.

In the antiperovskite manganese nitrides Mn$_3A$N (where $A$~=~Ni, Ga, Sn, etc.), the Mn atoms form a kagome lattice in the (111) plane. 
The noncollinear AFM Mn$_3A$N with a nonzero Berry curvature has been predicted to exhibit a large anomalous Hall effect (AHE) and an anomalous Nernst effect even with a quite small canted magnetization of the order of $0.001$--$0.01$~$\mu_{\rm B}$ per atom~\cite{Gurung MGN, Mn3AN_2, Mn3AN_3}.
The AHE has been established in Mn$_3$Ni$_{1-x}$Cu$_x$N films~\cite{Cu-MNN PRB, Cu-MNN JAP}, strained Mn$_3$NiN films~\cite{Boldrin MNN}, and strained Mn$_3$SnN films~\cite{Mn3SnN APL}.
Although electrical current switching of both Hall resistance with no external magnetic field and nonlinear Hall resistance with respect to an external magnetic field have been reported in MGN/Pt bilayers, no clear evidence of AHE has yet been presented~\cite{Hajiri APL}.
In addition, a thermal contribution due to pulse current injection such as a thermal activation effect, a joule heating effect, or an electromigration effect can change the Hall resistance~\cite{electromigration PRL, electromigration JAP, Meinert thermal activation}.
Hence, as has been pointed out, both the low-electrical-current writing and reading operations of MGN/Pt bilayers could be of nonmagnetic thermal origin~\cite{Adv. Mater.}.

\begin{figure*}[t]
\includegraphics[width=\linewidth,clip]{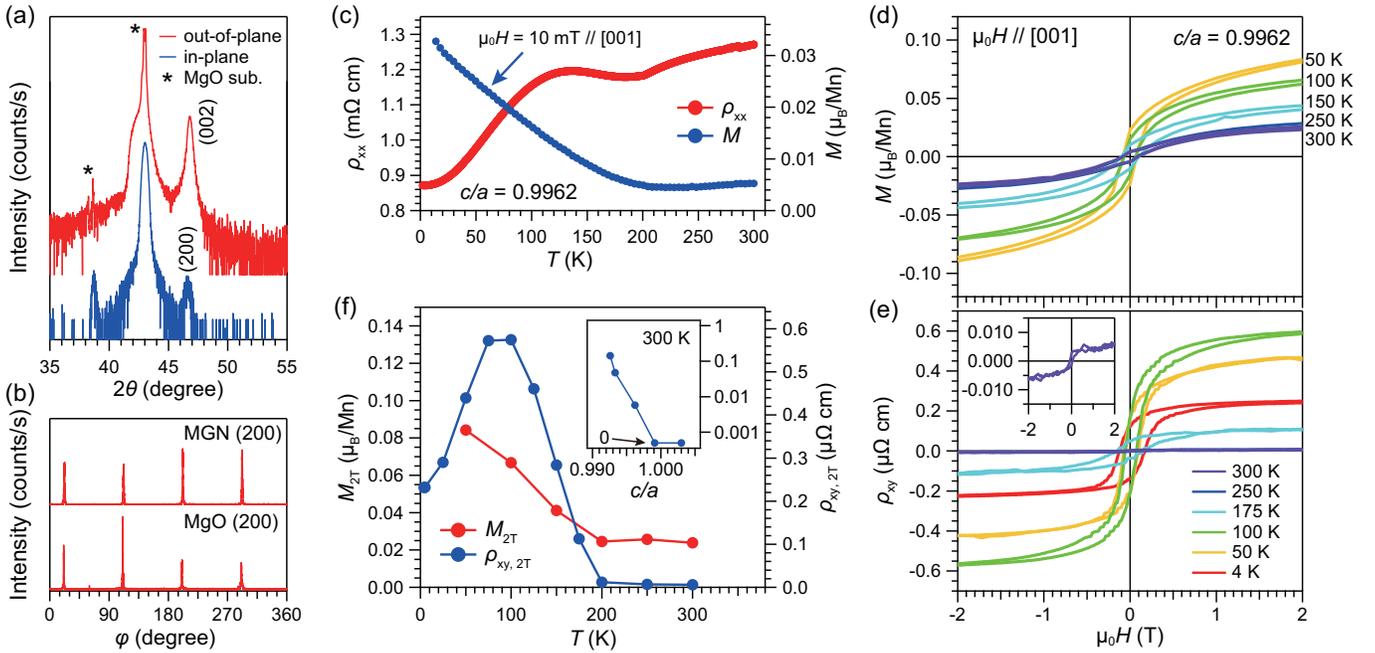}
\caption{
(a) Out-of-plane and in-plane XRD profiles for MGN films around (002)/(200) peaks.
(b) In-plane $\phi$ scans of MGN films.
(c) Temperature ($T$) dependence of the resistivity $\rho_{\mathrm{xx}}$ and magnetization $M$ of MGN films. 
For the magnetization measurement, an external magnetic field $\mu_{0}H$ of 10~mT was applied parallel to the [001] direction.
(d) Out-of-plane magnetic hysteresis loops and (e) anomalous Hall resistivity loops for MGN films at various temperatures.
Inset of panel~(e) is an enlargement of the  anomalous Hall resistivity loop at 300~K. 
(f) Out-of-plane magnetization at 2~T and anomalous Hall resistivity $\rho_{\mathrm{xy},\rm{2T}}$ at 2~T for MGN films as a function of temperature.
The inset presents the anomalous Hall resistivity at 300~K as a function of the $c/a$ ratio.
$c/a=0.9962$ film is 35~nm thick, while other films are 50~nm thick.
}
\label{fig:one}
\end{figure*}

In this investigation, we studied the magnetotransport properties of MGN films and performed systematic switching operations of strained MGN/HM (HM~=~Pt, Ta) bilayers.
At room temperature, no AHE was obtained in relaxed MGN films, whereas the AHE appeared with respect to the ratio between the lattice constants $c$ and $a$ ($c/a$ ratio).
The temperature dependence of the magnetization and the AHE suggests that skew scattering is dominant below 200~K, whereas another origin, probably noncollinear AFM order with nonzero Berry curvature, is dominant above 200~K.
To estimate the effect of thermal contributions on write/read operations, the pulse-width dependence and relaxation after pulse injection were measured.
We show that the thermal activation effect, joule heating effect, and electromigration effect play a minor role.
The existence of multistate signal amplitude with respect to pulse current with no external magnetic field was demonstrated.

\section{EXPERIMENTAL DETAILS}
MGN films were grown by reactive magnetron sputtering on MgO(001) substrates using a Mn$_{3}$Ga target at 400~$^\circ$C.
As details of the film growth were reported in our previous work, the $c/a$ ratio was controlled by precise control of N$_2$ partial pressure during film growth~\cite{Ishino AIP advances}.
The crystal structure was analyzed using both in-plane and out-of-plane X-ray diffraction (XRD) measurements with Cu~$K\alpha$ radiation.
Magnetic properties were characterized using superconducting quantum interference device magnetometry. 
Transport properties were characterized by the standard DC four-terminal method.
For write/read operations, a layer of Pt or Ta (3nm) was deposited by magnetron sputtering at room temperature after film growth.
20~$\mu$m-width Hall bars with Ti/Cu contact pads were prepared by a conventional photolithographic process. 
All SOT measurements were performed with no external magnetic field at room temperature (300~K).
The sequence of write/read operations was the same as described in our previous report~\cite{Hajiri APL}.

\section{RESULTS AND DISCUSSIONS}
\subsection{Film Characteristics and Magnetotransport Properties}
Typical out-of-plane $2\theta$--$\omega$ and in-plane $2\theta_{\chi}$--$\phi$ X-ray diffraction (XRD) patterns for the 35-nm-thick MGN films are shown in Fig.~\ref{fig:one}(a). 
Only the MGN (002) and (200) planes exhibit Bragg peaks, in the out-of-plane and in-plane XRD patterns, respectively. 
In addition, epitaxial growth is confirmed by the results of $\phi$-scan measurement as shown in Fig.~\ref{fig:one}(b), showing that their epitaxial relationship is MgO(001)[100]//MGN(001)[100]. 
The lattice constants $c$ and $a$ are 0.38817~nm and 0.38965~nm, respectively, giving $c/a=0.9962$.
The temperature dependencies of the resistivity $\rho_{\mathrm{xx}}$ and magnetization $M$ are shown in Fig.~\ref{fig:one}(c).
In the $\rho_{\mathrm{xx}}$ curve, there is a clear anomaly at 200~K.
Likewise, an FM-like transition is observed at 200~K in the magnetization curve. 
These temperature dependencies were also observed in our previous switching study of MGN/Pt films~\cite{Hajiri APL}.
The ground state of MGN at room temperature is well known to exhibit $\Gamma_{5g}$ spin structure~\cite{MGN neutron 1987}.
The coexistence of $\Gamma_{5g}$ and M-1 phase below the FM-like transition temperature has also been previously reported~\cite{MGN neutron 2016}.
As the MGN/FM bilayers exhibit an exchange bias at 4~K~\cite{MGN exchange bias}, it is concluded that $\Gamma_{5g}$ order and M-1 phase coexist below 200~K.

Figure~\ref{fig:one}(d) and \ref{fig:one}(e) show the magnetic $M$ hysteresis and anomalous Hall resistivity $\rho_{\mathrm{xy}}$ loops for the MGN films measured along the $\langle001\rangle$ direction at various temperatures.
In the $\rho_{\mathrm{xy}}$ loops, linear contribution from the ordinary Hall effect has been subtracted.
The $\rho_{\mathrm{xy}}$ loops before subtract the ordinary Hall effect and a summary of the Hall coefficient values are given in Fig.~\ref{fig:six} of Appendix~A.
In both loops, hysteresis is clearly exhibited not only in the coexistence phase below 200~K but also in the $\Gamma_{5g}$ single phase.
Both the magnetization and the $\rho_{\mathrm{xy}}$ loops of the MGN films have similar  coercive field ($H_c$) values, indicating that the AHE is directly related to the MGN magnetic order.
The temperature dependencies of the magnetization and $\rho_{\mathrm{xy}}$ at 2~T are shown in Fig.~\ref{fig:one}(f).
Above 200~K, the magnetization remains nearly constant at $\sim0.02$~$\mu_{\rm B}/\mathrm{Mn}$, and below 200~K, it increases monotonically with decreasing temperature.
$\rho_{\mathrm{xy}}$, on the other hand, increases slightly with decreasing temperature above 200~K and increases dramatically with decreasing temperature below 200~K, indicating that the increase in  $\rho_{\mathrm{xy}}$ is strongly linked to the increase in magnetization.
In contrast to the magnetization, however, $\rho_{\mathrm{xy}}$ begins to decrease below 75~K.
As $\rho_{\mathrm{xy}}$ is found to be proportional to $\rho_{\mathrm{xx}}$ below 75~K, $\rho_{\mathrm{xy}}$ below 200~K would be dominated by a net magnetization.

According to theoretical studies on piezomagnetism of MGN with $\Gamma_{5g}$ order, a net magnetization can appear by the induction of strain~\cite{Lukashev PRB, Zeman MGN}, and the magnitude of the net magnetization increases linearly with respect to the strain $\epsilon$~\%, with a coefficient of 0.013~$\mu_{\rm B}/\mathrm{Mn}/\%$~\cite{Lukashev PRB}.
Our MGN films show $\epsilon$ and remnant magnetization of approximately 0.3~\% and 0.004~$\mu_{\rm B}/\mathrm{Mn}$ at 300~K, respectively.
As the net magnetization observed in our MGN films is similar to the value calculated theoretically, it is considered that the canted $\Gamma_{5g}$ order is realized above 200~K.
The $\Gamma_{5g}$ order does not show the AHE because it has mirror symmetry, and the symmetry operations make the Berry curvature vanish after integration over the entire Brillouin zone~\cite{Gurung MGN}.
In antiperovskite nitride films with $\Gamma_{5g}$ order, however, the anomalous Hall conductivity (AHC) tensor is reported to be highly sensitive to strain.
Although strain-free Mn$_3$NiN films show no AHE~\cite{Cu-MNN PRB}, strained Mn$_3$NiN films do show AHE~\cite{Boldrin MNN}.
This is explained by the reduction of the symmetry of a space group, under which nonzero Berry curvature is induced when a finite strain is applied~\cite{Boldrin MNN}.
Strained Mn$_3$SnN films likewise show a large AHE, but it is suggested that the biaxial strain induces $\Gamma_{4g}$ order from $\Gamma_{5g}$ order~\cite{Mn3SnN APL}.
These past findings indicate that AHE cannot be accounted for by either extrinsic scattering processes or changes in magnetization; hence, nonzero Berry curvature plays an important role.
In our MGN films, $\rho_{\mathrm{xy}}$ is observed to increase with decreasing $c/a$ ratio at 300~K as shown in the inset of Fig.~\ref{fig:one}(f), highlighting the fact that appearance of AHE at 300~K in MGN films is also strongly related to the film strain and/or reduced magnetic space group.
Therefore, although we cannot experimentally separate the contributions from canted net magnetization and nonzero Berry curvature due to noncollinear AFM order above 200~K, we can state that part of the AHE comes from noncollinear AFM order.
From the view point of SOT, we will discuss the possible origin of AHE in the later section.

\subsection{Dependence of Reversible Electrical Switching on HM}

\begin{table}[b]
\begin{center}
\caption{
Resistivities of Mn$_3$GaN, Pt, and Ta single-layer films, and Mn$_3$GaN/Pt and Mn$_3$GaN/Ta bilayers at 300~K.
The resistivities of Mn$_3$GaN/Pt and Mn$_3$GaN/Ta bilayers were calculated using a parallel circuit model from the corresponding single layer resistivities.
}
\renewcommand{\arraystretch}{1.5}
\begin{tabular}{m{9em}m{12em}c}
\hline
Film  & $\rho_{\mathrm{xx}}$ ($\mu\Omega$~cm) \\ 
\hline
Mn$_3$GaN  & 1271 \rule[0mm]{0mm}{4mm}\\
Pt  & 112 \rule[0mm]{0mm}{4mm}\\
Ta  & 972\rule[0mm]{0mm}{4mm}\\

Mn$_3$GaN/Pt  & 504~~(exp.),~~541~(calc.)\rule[0mm]{0mm}{4mm}\\
Mn$_3$GaN/Ta  & 1261~(exp.),~~1222~(calc.)\rule[0mm]{0mm}{4mm}\\
\hline
\end{tabular} 
\label{table:one}
\end{center}
\end{table}

\begin{figure}[t]
\begin{center}
\includegraphics[width=\linewidth,clip]{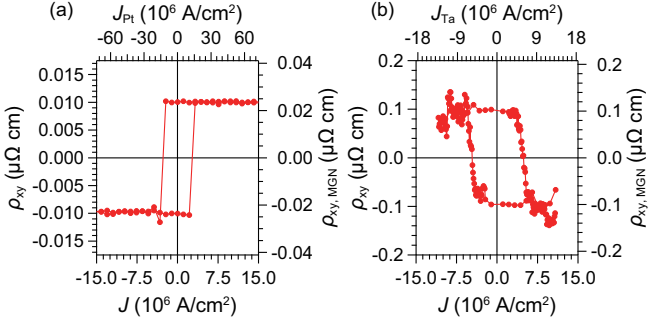}
\caption{
Sequential electrical current switching for (a) MGN/Pt and (b) MGN/Ta bilayers with no external magnetic field at room temperature.
The current density $J$ is derived by the total thickness of the bilayers, and the current density through HM layer $J_{HM}$ is derived by HM thickness and the current through the HM layer as estimated using a parallel circuit model.
In the same manner, $\rho_{\mathrm{xy}}$ and $\rho_{xy,\rm{MGN}}$ are derived by the total thickness of bilayers and by MGN thickness and the current through MGN layer.
}
\label{fig:two}
\end{center}
\end{figure}

For the electrical write/read operations, typical Hall bar devices of bilayers consisting of strained-MGN~(20~nm) with either Pt or Ta~(3~nm), with 20~$\mu$m width were prepared.
From a parallel circuit model, the $\rho_{\mathrm{xx}}$ values for both the MGN/Pt and MGN/Ta bilayers can be derived using $\rho_{\mathrm{xx}}$ of each single-layer film, which enable us to estimate  current density through the Pt and Ta layers ($J_{\mathrm{c}, HM}$).
The $\rho_{\mathrm{xx}}$ values at 300~K for MGN, Pt, and Ta single-layer films and MGN/Pt and MGN/Ta bilayers are given in Table.~\ref{table:one}.
Using these bilayers, sequential write/read operations were performed, with the results as shown in Fig.~\ref{fig:two}.
Here, the top and bottom axes are the current densities derived from the Pt/Ta layer alone and the bilayer total thickness, respectively, and the left and right axes are the $\rho_{\mathrm{xy}}$ values derived from the full bilayer and the MGN layer alone, respectively.
In addition, the constant offset probably due to geometrical imperfections of the Hall bar and/or thermoelectric voltage was subtracted.
$\rho_{\mathrm{xy}}$ changes with respect to $J$, and a clear hysteresis loop is observed in both bilayers at 300~K with no external magnetic field, indicating success of the electrical write/read operation.
The critical current densities $J_\mathrm{c}$ ($J_{\mathrm{c}, HM}$) for MGN/Pt and MGN/Ta were obtained as $2.72\times10^6$~A/cm$^{2}$ ($13.13\times10^6$~A/cm$^{2}$) and $4.81\times10^6$~A/cm$^{2}$ ($6.13\times10^6$~A/cm$^{2}$), respectively.
Theoretically, $J_\mathrm{c}$ would be proportional to $\theta_{HM}^{-1}$, where $\theta_{HM}$ is the spin Hall angle of the HM layer~\cite{Yamane PRB}.
We find that the ratio $J_{\mathrm{c}, \rm{Pt}}:J_{\mathrm{c}, \rm{Ta}}=2.1:1$ is nearly equal to the ratio $|\theta_{\rm{Pt}}|^{-1}:|\theta_{\rm{Ta}}|^{-1}=1.9:1$, indicating that spin current generated in the HM layer plays an important role.
We note that the different $\rho_{xy}$ switching width between MGN/Pt and MGN/Ta devices is probably due to local variations in the quality and/or strain of the thin films, which is discussed in Fig.~\ref{fig:seven} of Appendix~B.
Besides, the comparison of $\rho_{xy}$ width between field-sweep and SOT measurements using the same device is presented in Fig.~\ref{fig:eight} of Appendix~C.

\begin{figure}
\begin{center}
\includegraphics[width=\linewidth,clip]{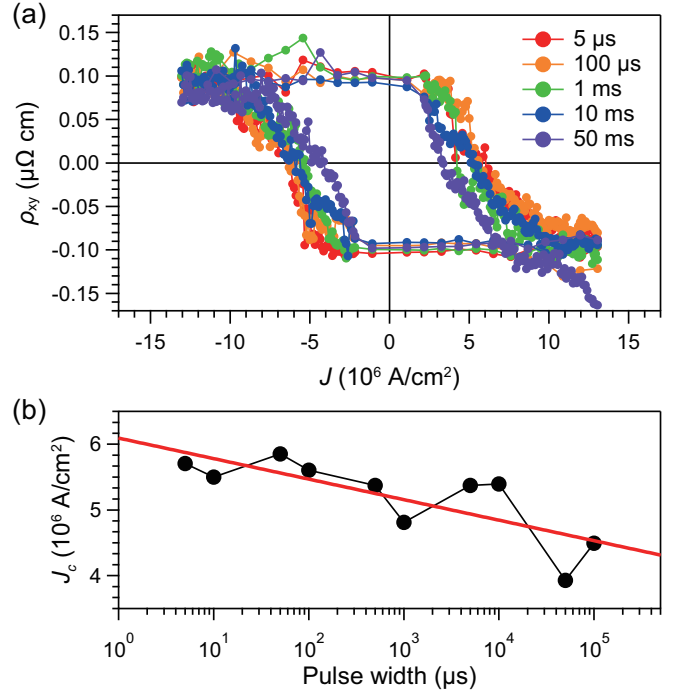}
\caption{
(a) $\rho_{\mathrm{xy}}$ as a function of $J$ observed for several pulse widths in MGN/Ta bilayers with no external magnetic field at room temperature.
(b) Critical current density $J_\mathrm{c}$ as a function of pulse width. 
The bold solid line is the result of a fit to a thermal activation model. 
}
\label{fig:three}
\end{center}
\end{figure}

To evaluate the effect of joule heating by applying pulse current, the pulse-width dependence was investigated.
Figure~\ref{fig:three}(a) shows $\rho_{\mathrm{xy}}$ as a function of $J$ with several pulse widths for MGN/Ta bilayers.
Although the switching $\rho_{\mathrm{xy}}$ amplitudes remain nearly the same, the hysteresis loops become slightly narrower with increasing pulse width.
$J_\mathrm{c}$ is plotted as a function of pulse width in Fig.~\ref{fig:three}(b).
Here, $J_\mathrm{c}$ is fitted by the thermal activation model~\cite{thermal activation model};

\begin{equation}
J_\mathrm{c}=J_{\mathrm{c}0}\left[1-\frac{1}{\Delta}\rm{ln}(\frac{\tau}{\tau_0})\right],
\end{equation}

where $J_{\mathrm{c}0}$ is the critical current density at 0~K, $\Delta$ is the thermal stability factor, and $\tau_0^{-1}$ is the thermally activated switching frequency, for which we assume a frequency of $1/\tau_0=1$~THz.
It can be seen that $J_\mathrm{c}$ can be fitted by the thermal activation model with $J_{\mathrm{c}0}=7.97\times10^6$~A/cm$^{2}$ and $\Delta=58.7$.
These results highlight the finding that a $J_\mathrm{c}$ value of the order $10^6$~A/cm$^{2}$ originates intrinsically in spin torque whereas the thermal activation, through it exists, plays a minor role.

\subsection{Thermal Contribution to Write/Read Operations}
\begin{figure}[t]
\begin{center}
\includegraphics[width=\linewidth,clip]{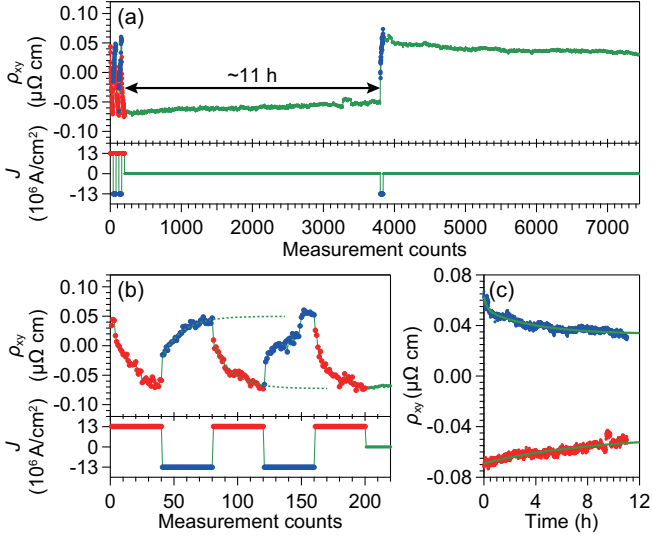}
\caption{
(a) Continuous write/read operations and subsequent relaxation measurements with no external magnetic field at room temperature.
(b) Continuous write/read operation part from panel~(a).
The dashed lines are the results of a fit to the exponential decay function.
(c) Relaxation measurement parts from panel~(a).
The bold solid lines are the results of a fit to the double exponential function.
}
\label{fig:four}
\end{center}
\end{figure}

Because heating and electromigration effects can affect $\rho_{\mathrm{xy}}$ signal amplitudes~\cite{electromigration JAP}, the relaxation behavior after switching was investigated.
Figure~\ref{fig:four} presents the continuous write/read operation using $I_{\rm{pulse}}=\pm13\times10^6$~A/cm$^{2}$ with a 5~$\mu$s pulse width and subsequent relaxation measurements for MGN/Ta bilayers at 300~K.
Figure~\ref{fig:four}(b) is an enlargement of the continuous $\pm I_{\rm{pulse}}$ write part shown in Fig.~\ref{fig:four}(a). 
As discussed for synthetic AFM materials~\cite{Moriyama PRL} and in our previous MGN/Pt study~\cite{Hajiri APL}, the observed asymptotic $\rho_{\mathrm{xy}}$ behavior can be fitted by an exponential decay function $y=y_0+A_{~}\mathrm{exp}\left[-(x+x_0)/\tau\right]$ with time constants $\tau$ of 14.9 pulse number (197.0~s) for $+I_{\rm{pulse}}$ and 15.7 pulse number (207.6~s) for $-I_{\rm{pulse}}$ with continuous write/read operations in approximately 13.25~s cycles.
The relaxation behavior after switching was measured after three cycles of the continuous $\pm I_{\rm{pulse}}$ write operations.

Figure~\ref{fig:four}(c) presents the $\rho_{\mathrm{xy}}$ relaxation as a function of time after continuous $\pm I_{\rm{pulse}}$ write operations.
The relaxation behavior is characterized by fit to a double exponential function~\cite{electromigration JAP},

\begin{equation}
d=d_0+d_1~\mathrm{exp}\left(-\frac{t}{\tau_1}\right)+d_2~\mathrm{exp}\left(-\frac{t}{\tau_2}\right),
\end{equation}

where $d_0$ is a base line and is the value reached when attenuated, $d_{1}$, $d_{2}$ are amplitude parameters, and $\tau_{1}$, $\tau_{2}$ are the relaxation times.
Both relaxations after $\pm I_{\rm{pulse}}$ write operations fit well with $d_0=-0.0467$~$\mu\Omega$cm, $d_1=-0.0219$~$\mu\Omega$cm, $d_2=-0.0003$~$\mu\Omega$cm, $\tau_1=8.63$~h, $\tau_2=11.67$~h for after the $+I_{\rm{pulse}}$ write operation, and $d_0=0.0331$~$\mu\Omega$cm, $d_1=0.0100$~$\mu\Omega$cm, $d_2=0.0194$~$\mu\Omega$cm, $\tau_1=0.25$~h, $\tau_2=3.89$~h for after the $-I_{\rm{pulse}}$ write operation.
With regard to the change in Hall resistance due to the effects of annealing and electromigration, the short and long decay times are reported to be approximately 4~min and 50~min, respectively~\cite{electromigration JAP}.
Compared with these values, both the short and long decay times observed here are five to ten times longer.
In addition, for the change in Hall resistance due to annealing and electromigration effects, the Hall resistance asymptotes to the initial value before pulse injection is a few minutes to a few hours~\cite{electromigration JAP, Meinert MnN}, in contrast with our MGN/Ta bilayers, for which $d_0$ is a rather large value.
Whereas our electrical measurements do not enable us to fully distinguish SOT and other possible contributions such as thermal activation and electromigration effects, the relaxation behavior and pulse-width behavior in the MGN/Ta bilayers highlight the fact that SOT plays an important role in the present switching behavior.
In the case of collinear AFM materials such as NiO, $90^\circ$ switching of the N$\rm{\acute{e}}$el vector is required, which can be achieved by the flow of current in two orthogonal directions.
However, current flow in orthogonal directions causes inhomogeneous current density due to the current crowding effect, which induces a change in Hall resistance due to annealing and electromigration effect~\cite{electromigration JAP}.
In the case of noncollinear AFM materials, in contrast, $180^\circ$ switching of each noncollinear spin is required, which can be achieved by the flow of current in a straight line; this implies that the current crowding effect in this study can be considered small.

\begin{figure}[t]
\begin{center}
\includegraphics[width=\linewidth,clip]{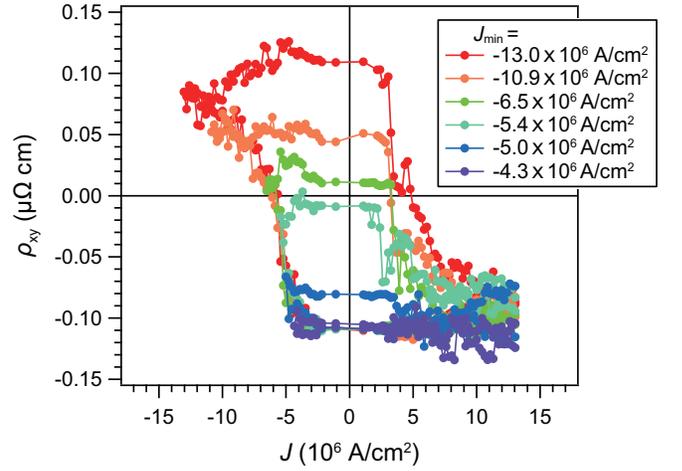}
\caption{
Dependence of $\rho_{\mathrm{xy}}$--$J$ loops on $J_{\rm{min}}$ for MGN/Ta bilayers with no external magnetic field at room temperature.
}
\label{fig:five}
\end{center}
\end{figure}

\subsection{Memristive Switching with Respect to Pulse Current Density}
Finally, here we discuss the memristive behavior of MGN/Ta bilayers.
Figure~\ref{fig:five} shows $\rho_{\mathrm{xy}}$--$J$ loops at 300~K, where $J_{\rm{max}}=13.0\times10^6$~A/cm$^{2}$ was first applied and subsequently scanned from $J_{\rm{max}}$ to $J_{\rm{min}}$ and from $J_{\rm{min}}$ to $J_{\rm{max}}$.
$\rho_{\mathrm{xy}}$ exhibits multiple stable signal amplitudes according to the magnitude of $J_{\rm{min}}$.
The same behavior has been observed in Mn$_3$Sn/Pt bilayers~\cite{Mn3Sn SOT} and AFM/FM bilayers~\cite{Fukami AFM/FM SOT}, suggesting that the phenomenon of multiple stable magnitudes of the Hall resistance originates in the multi-AFM domain character, which allows us to tune the signal amplitude in an analog manner. 
In contrast to Mn$_3$Sn/Pt bilayers~\cite{Mn3Sn SOT}, MGN/Ta bilayers show a memristive behavior with no external magnetic field, highlighting the advantage of Mn$_3A$N systems for use in neuromorphic computing.

According to theoretical study of SOT of $\Gamma_{4g}$ order, the noncollinear spins rotate in the (111) plane of the kagome lattice, where the injected spins are directed perpendicular to the kagome lattice.
Therefore, $J_\mathrm{c}$ is determined by an in-plane anisotropic energy of the kagome lattice and the injected spin direction.
Further improvements in SOT efficiency in noncollinear Mn$_3A$N systems may be attained using (110)-oriented films of low-anisotropy materials.
Theoretically, Mn$_3$Ga$_{1-x}$Ni$_x$N~\cite{MGN SOT theory} has a low anisotropy energy, implying that it would be worthwhile to investigate the dependence of SOT efficiency on $A$ atoms.
On the other hand, if the AHE comes from only net magnetization (for simplicity, assume a net magnetization in the perpendicular direction like Ferrimagnet), the magnetic field parallel to the current direction is generally needed to realize the SOT switching. 
In contrast, SOT of noncollinear AFM theoretically satisfies even though with no external magnetic field~\cite{Yamane PRB, MGN SOT theory}. 
Indeed, no switching has been observed with no external magnetic field at low temperatures where the Ferrimagnetic M1 phase is dominant in the AHE~\cite{Hajiri APL}. 
Although the joule heating can affect to the Hall resistivity, the pulse-width measurement and relaxation measurement show a heating effect plays a minor role. 
From these results, we can conjecture that AHE is possibly related to noncollinear AFM order from the view point of SOT at 300 K.

\section{Conclusion}
In this study, we have shown the AHE and SOT switching of noncollinear AFM MGN at room temperature.
By tuning the $c/a$ ratio, two origins of AHE were observed: one for AHE above 200~K, possibly related to noncollinear AFM order, and one for AHE below 200~K, dominated by magnetization.
Using MGN/HM bilayers, we have demonstrated the SOT switching of noncollinear AFM spin in MGN at room temperature with no external magnetic field. 
The effects of thermal activation on $J_\mathrm{c}$ and the effects of heating and electromigration on $\rho_{\mathrm{xy}}$ were excluded by pulse-width measurements and relaxation measurements after pulse injection.
In addition, multistate memristive switching with respect to pulse current density was demonstrated.
These results show that, efficient SOT can be attained in MGN/HM bilayers with memristive functionality with no external magnetic field, and demonstrate the potential application in AFM spintronics and neuromorphic computing.

\begin{acknowledgments}
This work was supported by the Japan Society for the Promotion of Science (KAKENHI Grant Nos. 20H02602 and 19K15445), Tokuyama Science Foundation, the Hori Science and Arts Foundation, and Kyosho Hatta Foundation. 
Part of this work was carried out under the Cooperative Research Project Program of the Research Institute of Electrical Communication, Tohoku University.
\end{acknowledgments}

\section{Appendix}
\appendix
\section{Ordinary Hall effect of MGN films}
\begin{figure}[h]
\begin{center}
\includegraphics[width=0.9\linewidth,clip]{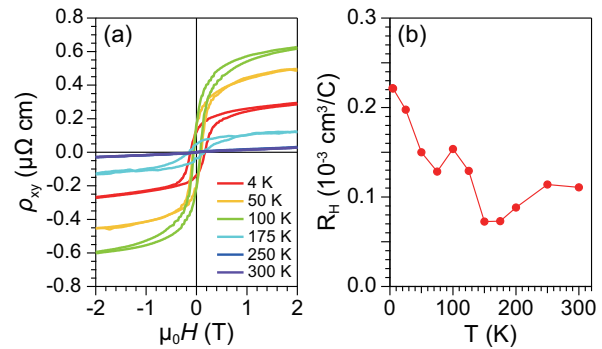}
\caption{
(a) Hall resistivity $\rho_{xy}$ as a function of external magnetic field for MGN films ($c/a=0.9962$) before subtract the ordinary Hall effect.
(b) Temperature dependence of the Hall coefficient $R_H$.
}
\label{fig:six}
\end{center}
\end{figure}

Figure~\ref{fig:six}(a) shows the Hall resistivity $\rho_{xy}$ as a function of external magnetic field for MGN films ($c/a=0.9962$) at various temperature before subtract the ordinary Hall effect.
The temperature dependence of the Hall coefficient $R_H$ is summarized in Fig.~\ref{fig:six}(b).

\section{Sample dependence of $\rho_{xy}$ switching width in SOT measurements}

\begin{figure}[]
\begin{center}
\includegraphics[width=0.7\linewidth,clip]{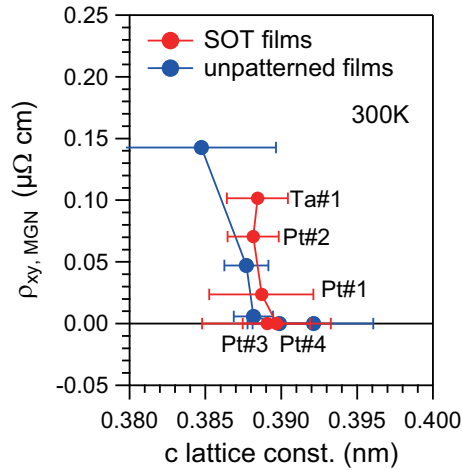}
\caption{
Sample dependence of $\rho_{xy}$ switching width in SOT measurements for MGN(20~nm)/HM(3~bn) bilayers.
MGN/Pt and MGN/Ta results presented in the main text are the devices Pt\#1 and Ta\#1, respectively.
For Pt\#2 device, $\rho_{xy, \rm{MGN}}$ was derived by assuming that the current flows through MGN and Pt in the same proportion as in Pt\#1 device.
}
\label{fig:seven}
\end{center}
\end{figure}

Figure~\ref{fig:seven} shows the sample dependence of $\rho_{xy}$ switching width in SOT measurements as a function of $c$ lattice constant.
The $\rho_{xy}$ obtained by the field sweep measurements of unpatterned films is also plotted.
While no $\rho_{xy}$ switching is observed for samples with $c$ lattice constants longer than 0.3890~nm, $\rho_{xy}$ switching is observed in samples with shorter $c$ lattice constants, which is consistent with the results of AHE measurements of unpartterned films.
On the other hand, although no large difference in $c$ lattice constant between 0.3882 and 0.3887~nm among SOT films in which $\rho_{xy}$ switching were observed, $\rho_{xy}$ switching width shows sample dependence.
Since the amplitude of $\rho_{xy}$ is strongly related to $c/a$ ratio as discussed in the main text, the sample dependence of $\rho_{xy}$ switching width probably due to local variations in the quality and/or strain of the thin films.

\section{Comparison of $\rho_{xy}$ width between field-sweep and SOT measurements using same device}

\begin{figure}[t]
\begin{center}
\includegraphics[width=\linewidth,clip]{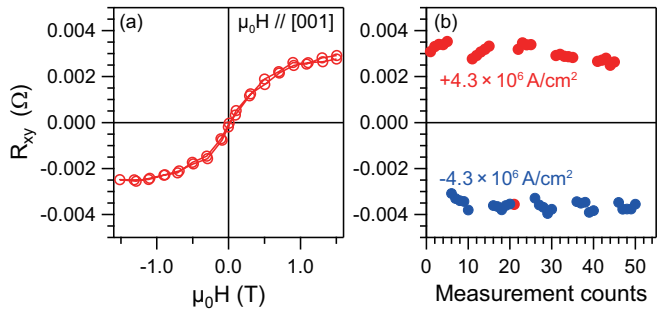}
\caption{
Hall and SOT measurements using the same SOT device.
(a) $R_{xy}$ after subtract estimated ordinary Hall effect vs external magnetic field of MGN/Pt bilayers.
(b) $R_{xy}$ vs pulse number of MGN/Pt bilayers. 
All measurements were performed at room temperature using the same MGN/Pt Hall device.
The data are reproduced from $Appl. Phys. Lett.$ {\bf 115}, 052403 (2019)~\cite{Hajiri APL}, with the permission of AIP Publishing.
}
\label{fig:eight}
\end{center}
\end{figure}

Figure~\ref{fig:eight} shows the results of the Hall and SOT measurements using the same SOT device.
Unfortunately, since our superconducting magnet has limited bore sizes, we cannot measure AHE of SOT devices. 
On the other hand, as shown in Fig.~\ref{fig:eight}(a), the nonlinear Hall effect was observed within $\pm1.5$~T range in SOT devices~\cite{Hajiri APL}. 
Although the hysteresis was not observed probably due to insufficient magnetic field, we confirmed that the change of Hall resistance $R_{xy}$ is similar between field sweep and SOT measurements in the same sample as shown in Fig.~\ref{fig:eight}(b).

\end{document}